\documentclass[aps,prl,twocolumn,groupedaddress]{revtex4}
\usepackage{graphicx} 
\usepackage{dcolumn}% Align table columns on decimal point
\usepackage{bm}% bold math
\begin{document}

\title{Suppression of the N\'eel temperature in hydrothermally synthesized $\alpha$-Fe$_{2}$O$_{3}$
nanoparticles} 
\author{Jun Wang$^{1,a}$, Wei Wu$^{1}$, Fan Zhao$^{1}$, and Guo-meng Zhao$^{1,2,b}$} 
\affiliation{
$^{1}$Department of Physics, Faculty of Science, Ningbo
University, Ningbo, P. R. China~\\
$^{2}$Department of Physics and Astronomy, 
California State University, Los Angeles, CA 90032, USA}

\begin{abstract}

 Magnetic measurements up to 1000 K have been performed on hydrothermally 
 synthesized $\alpha$-Fe$_{2}$O$_{3}$ nanoparticles (60 nm) using a Quantum
 Design vibrating sample magnetometer. A high vacuum environment
 (1$\times$10$^{-5}$ torr) during the magnetic measurement up to 1000 K leads to a complete 
 reduction of $\alpha$-Fe$_{2}$O$_{3}$  to Fe$_{3}$O$_{4}$. This
 precludes the determination of the N\'eel temperature for the $\alpha$-Fe$_{2}$O$_{3}$
 nanoparticles. In contrast, coating
 $\alpha$-Fe$_{2}$O$_{3}$ nanoparticles with SiO$_{2}$ stabilizes the $\alpha$-Fe$_{2}$O$_{3}$
 phase up to 930 K, which allows us to determine the N\'eel temperature of the $\alpha$-Fe$_{2}$O$_{3}$
 nanoparticles for the first time. The N\'eel temperature of the 60-nm 
 $\alpha$-Fe$_{2}$O$_{3}$ nanoparticles is found to be 945 K, about 15 K below the 
 bulk value. The small reduction of the N\'eel temperature of the $\alpha$-Fe$_{2}$O$_{3}$ 
 nanoparticles is consistent with a finite-size scaling theory. Our
 current results also show that coating nanoparticles with SiO$_{2}$
 can effectively protect nanoparticles from oxidation or
 reduction, which is important to technological applications.

\end{abstract}
\maketitle 

There are finite-size effects of ferromagnetism, antiferromagnetism,
superconductivity, and other physical properties when the dimensions
of solids are reduced to the nanoscale. 
For a magnetic system, the growth of the spin correlation length 
will be limited by the smallest dimension and the system will 
exhibit a reduced magnetic transition temperature following a finite-size scaling theory \cite{Fisher}.
 So far, experimental observations of finite-size effects in magnetic systems have
been mostly limited to quasi-two-dimensional ferromagnetic systems such as  
ferromagnetic thin films of Ni \cite{Li,Huang}, Fe \cite{Koon,Elmers},
Co \cite{Sch}, and Gd \cite{Far}. Experimental observation of finite-size
effects in antiferromagnetic (AF) thin films has been very challenging because it requires 
very thin layers and because the magnetic
response of an AF material is much weaker than that of a ferromagnet.
One study of finite-size effects
in an AF material is through the indirect measurement
of resistivity in thin layers of Cr, which is an incommensurate
spin-density-wave AF metal \cite{Full}. Since the
results were observed in Fe/Cr multilayers, the interlayer
coupling through the Cr layers and the ferromagnetic
ordering of the Fe layers complicate the analyses of
the finite-size scaling of Cr layers. The second study of finite-size effects 
in an AF material is through the direct magnetic measurement of  antiferromagnetic 
thin films of CoO that has an ideal N\'eel temperature ($T_{N}$) near
room temperature \cite{Amb95}. This study shows a finite-size scaling 
relationship with a scaling exponent $\nu$ of 0.65.

Recently, finite-size effects have been investigated in antiferromagnetic
hematite ($\alpha$-Fe$_{2}$O$_{3}$) nanowire arrays with a mean diameter $d$ of about 
150 nm \cite{Di}. A magnetic transition at 852 K has been assigned to the N\'eel 
temperature $T_{N}$ of the $\alpha$-Fe$_{2}$O$_{3}$
nanowire arrays \cite{Di}. Then the speculated $T_{N}$ of 852 K for the
nanowires is over 100 K below the bulk
value (960 K \cite{Di}). Such a $T_{N}$ reduction in these nanowires ($d$ = 150
nm) is too large to be compatible with the finite-size scaling
relations inferred from both ferromagnetic thin films \cite{Huang} and
antiferromagnetic thin films \cite{Amb95}.  Moreover, the 
assigned $T_{N}$ value \cite{Di} happens to be the same as the Curie temperature of 
Fe$_{3}$O$_{4}$. Therefore, it is likely that the magnetic
transition at 852 K is not associated with the antiferromagnetic
transition of the 150-nm $\alpha$-Fe$_{2}$O$_{3}$ nanowires but with
the minor phase of Fe$_{3}$O$_{4}$.

Here we report magnetic measurements (up to 1000 K) on hydrothermally 
 synthesized $\alpha$-Fe$_{2}$O$_{3}$ nanoparticles (60 nm) using a Quantum
 Design vibrating sample magnetometer. A high vacuum environment
 (1$\times$10$^{-5}$ torr) during the magnetic measurement up to 1000 K leads to a complete 
 reduction of $\alpha$-Fe$_{2}$O$_{3}$  to Fe$_{3}$O$_{4}$. This
 precludes the determination of the N\'eel temperature of the $\alpha$-Fe$_{2}$O$_{3}$
 nanoparticles. In contrast, coating
 $\alpha$-Fe$_{2}$O$_{3}$ nanoparticles with SiO$_{2}$ almost
 stabilizes the $\alpha$-Fe$_{2}$O$_{3}$
 phase up to 930 K, which allows us to determine the N\'eel temperature of the $\alpha$-Fe$_{2}$O$_{3}$
 nanoparticles for the first time. The N\'eel temperature of the 60-nm 
 $\alpha$-Fe$_{2}$O$_{3}$ nanoparticles is found to be 945 K, about 15 K below the 
 bulk value. The small reduction of the N\'eel temperature of the $\alpha$-Fe$_{2}$O$_{3}$ 
 nanoparticles is consistent with the finite-size scaling theory
 \cite{Fisher}. Our
 current results also show that coating nanoparticles with SiO$_{2}$
 can effectively protect nanoparticles from oxidation or
 reduction, which is important to technological applications of
 nanoparticles.

Samples of $\alpha$-Fe$_{2}$O$_{3}$ nanoparticles were synthesized by a 
hydrothermal route \cite{Hua}.  4 mL of 0.5 mol/L FeCl$_{3}$.6H$_{2}$O solution was
mixed with 12 mL of 0.5 mol/L CH$_{3}$COONa aqueous solution to form a mixture solution 
and 2 mL of C$_{2}$H$_{6}$O$_{2}$ was slowly added into the
mixture solution under vigorous stirring for 10 min. The whole
mixture was filled with 2 mL of H$_{2}$O and stirred for 30 min to form
homogeneous solution.  Then the solution was transferred into a Teflon lined
stainless steel autoclave, sealed, and maintained at 200 $^{\circ}$C for 12 h. After
the heating treatment, the autoclave was cooled to room temperature
naturally. Red product was obtained and collected by centrifugation,
washed with water several times, and finally dried at 60 $^{\circ}$C under
vacuum. The  $\alpha$-Fe$_{2}$O$_{3}$ nanoparticles were then coated
with SiO$_{2}$ according to the method of Ref.~\cite{Wang3}. Hematite particles were dispersed in 75 mL of 
ethanol, 15 mL of H$_{2}$O, and 10 mL of ammonia (25$\%$) by ultrasonic wave. The citric acid 
solutions (1 M) were added to the solution until flocculation was visible. The precipitate 
was redispersed by increasing the pH value with tetramethylammonium hydroxide. While stirring, 
0.2 mL of tetraethyl orthosilicate (TEOS) was slowly added to the ethanol mixture solution, and 
after 12 h another 0.3 mL of TEOS was added. The precipitate was collected by filtration and 
washed several times with distilled water and ethanol and finally dried in a vacuum 
oven at 60 $^{\circ}$C for about 6 h. 

\begin{figure}[htb]
     \vspace{-0.2cm}
    \includegraphics[height=6.5cm]{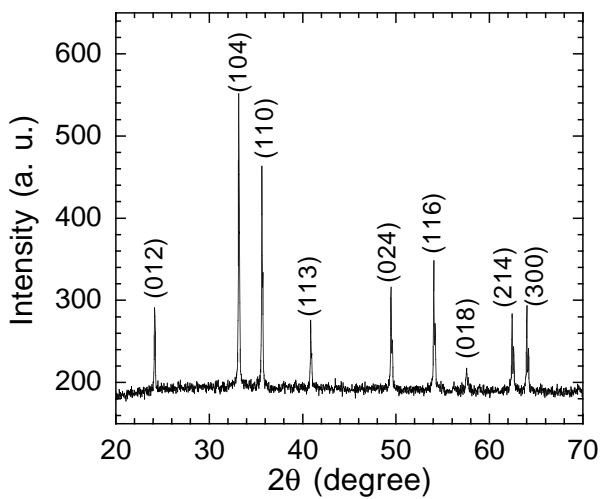}
     \vspace{-0.0cm}
 \caption[~]{a) X-ray diffraction (XRD) spectrum of  hydrothermally 
 synthesized $\alpha$-Fe$_{2}$O$_{3}$ nanoparticles. From the width of the
 (104) peak, the mean
 diameter of the sample is found to be 59.3 nm.} 
\end{figure}

Magnetization was measured using a Quantum Design vibrating sample magnetometer (VSM).
The moment measurement was carried out after the sample chamber reached a high vacuum of 
better than 9$\times$10$^{-6}$ torr. The absolute measurement
uncertainty 
in moment is less than 1$\times$10$^{-6}$
emu. The heating and cooling rates for
the magnetic measurements are 30 K/min.

Figure~1 shows x-ray diffraction (XRD) spectrum of  hydrothermally 
 synthesized $\alpha$-Fe$_{2}$O$_{3}$ nanoparticles. It is apparent that 
 the sample is of single phase. We can determine 
the average diameter from the width of the (104)
peak.  The full width 
at half 
maximum (HWHM) of 
the (104) peak is found to be
0.150$\pm$0.005$^{\circ}$ from a Gaussian fit. Using the Scherrer equation \cite{Klug}: 
$d = 0.89\lambda/(\beta \cos \theta)$ and with $\beta$ = 0.137$^{\circ}$
(after correcting for the instrumental broadening), we calculate 
$d$ = 59.3 nm. 

\begin{figure}[htb]
     \vspace{-0.2cm}
    \includegraphics[height=13cm]{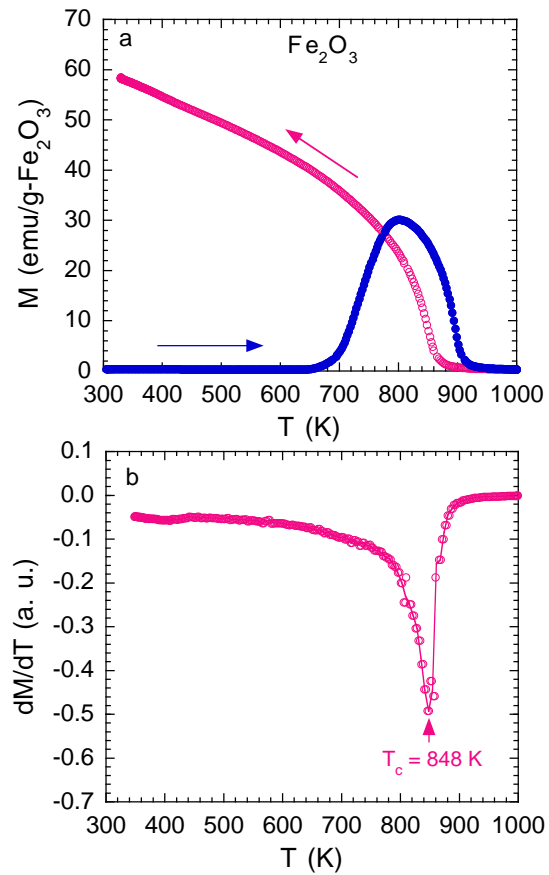}
     \vspace{-0.6cm}
 \caption[~]{a) Temperature dependence of the magnetization for a virgin
 sample of $\alpha$-Fe$_{2}$O$_{3}$ nanoparticles, which was measured 
 up to 1000 K. b) The derivative of the cooling-down magnetization ($dM/dT$) versus
 temperature for the sample. A minimum of $dM/dT$ at 848 K
 corresponds to a magnetic transition. } 
\end{figure}

Figure~2a shows magnetization versus temperature (up to 1000 K) for a virgin
 sample of $\alpha$-Fe$_{2}$O$_{3}$ nanoparticles, which 
 was measured in a magnetic field of
10 kOe.  One can clearly that, upon heating the magnetization shows a 
rapid rise above 650 K, which is the onset temperature of the
reduction of the weak-ferromagnetic $\alpha$-Fe$_{2}$O$_{3}$ to the ferrimagnetic
Fe$_{3}$O$_{4}$ phase. Upon cooling, the 
magnetization data show a magnetic transition at about 850 K (see
Fig.~2b), which is the 
same as the Curie temperature of Fe$_{3}$O$_{4}$. 

\begin{figure}[htb]
     \vspace{-0.2cm}
    \includegraphics[height=13cm]{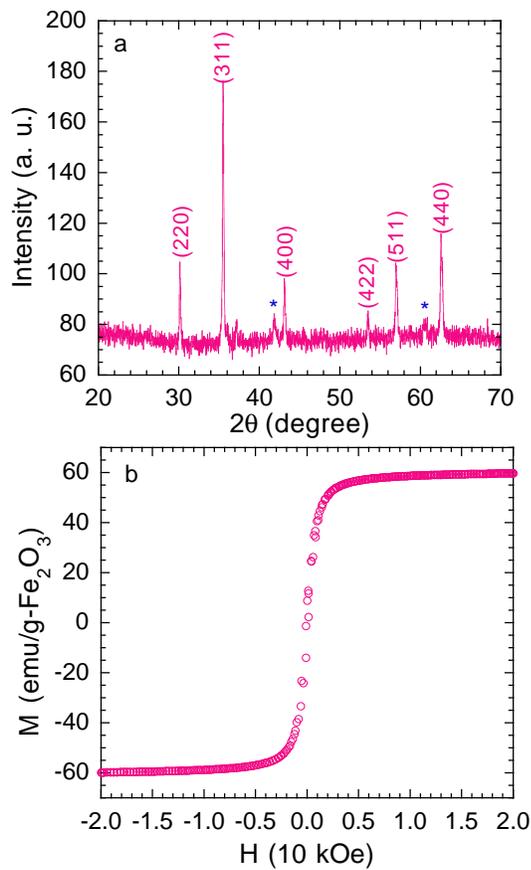}
     \vspace{-0.6cm}
 \caption[~]{a) X-ray diffraction spectrum of  
 the $\alpha$-Fe$_{2}$O$_{3}$ sample right after the magnetic
 measurement. All the XRD peaks can be indexed by the Fe$_{3}$O$_{4}$ 
 phase except for the peaks indicated by stars, which  respresent a minor phase of FeO
 (about 10$\%$). b) Magnetization versus magnetic field at 330 K after
 the sample was
 cooled down from 1000 K to 330 K.} 
\end{figure}

In order to check
whether all $\alpha$-Fe$_{2}$O$_{3}$ nanoparticles have been reduced
after the magnetic measurement up to 1000 K, we took x-ray diffraction
right after the magnetic measurement. The XRD spectrum is shown in
Fig.~3a. It is clear that the sample contains about 90$\%$ of Fe$_{3}$O$_{4}$ 
and about 10$\%$ of FeO. There is no trace of the $\alpha$-Fe$_{2}$O$_{3}$
phase, suggesting that the  $\alpha$-Fe$_{2}$O$_{3}$ nanoparticles have been 
completely reduced. Fig.~3b shows a magnetic hysteresis loop after
the sample was cooled down to 330 K.  From the loop, we determine the saturation 
magnetization $M_{s}$ to be 60 emu/g-Fe$_{2}$O$_{3}$.
After correction for about 10$\%$ of FeO, the saturation
magnetization of the converted  Fe$_{3}$O$_{4}$ phase is calculated to
be 68.3
emu/g-Fe$_{3}$O$_{4}$. Since the mean diameter of the converted  Fe$_{3}$O$_{4}$ phase
is about 45 nm, the  inferred $M_{s}$ for the 45-nm Fe$_{3}$O$_{4}$
nanoparticles is in quantitative agreement with the reported value
\cite{Goya}.

Figure~4 shows magnetization versus temperature (up to 930 K) for a
sample of $\alpha$-Fe$_{2}$O$_{3}$ nanoparticles coated with SiO$_{2}$. 
Upon heating, the magnetization starts to rise up above 690 K,
indicating reduction of $\alpha$-Fe$_{2}$O$_{3}$  to Fe$_{3}$O$_{4}$ nanoparticles.
The onset temperature of the reduction for the SiO$_{2}$-coated  $\alpha$-Fe$_{2}$O$_{3}$ nanoparticles
is about 40 K higher than that for the bare $\alpha$-Fe$_{2}$O$_{3}$ nanoparticles.
Upon cooling, the magnetization data show a magnetic transition at about 850 K which is the 
same as the Curie temperature of Fe$_{3}$O$_{4}$. The
difference between Fig.~2 and Fig.~4 is that the saturation
magnetization at 330 K for the SiO$_{2}$-coated  $\alpha$-Fe$_{2}$O$_{3}$ nanoparticles
is about 20$\%$ of that for the bare $\alpha$-Fe$_{2}$O$_{3}$ nanoparticles.
This implies that only about 20$\%$ of $\alpha$-Fe$_{2}$O$_{3}$ nanoparticles
have been reduced to Fe$_{3}$O$_{4}$ nanoparticles. Therefore, coating nanoparticles with SiO$_{2}$
can effectively protect the nanoparticles from oxidation or reduction, which is 
important to technological applications of
Ni, Fe, and Co nanoparticles where oxidation even occurs at room
temperature.

\begin{figure}[htb]
     \vspace{-0.2cm}
    \includegraphics[height=6.5cm]{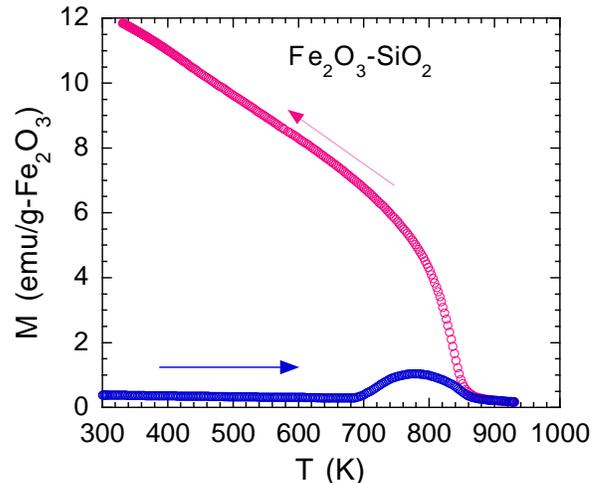}
     \vspace{-0.4cm}
 \caption[~]{Temperature dependence of the magnetization for
 for $\alpha$-Fe$_{2}$O$_{3}$ nanoparticles coated with SiO$_{2}$, which was measured 
 up to 930 K. Only about 20$\%$ of $\alpha$-Fe$_{2}$O$_{3}$ nanoparticles
have been reduced to Fe$_{3}$O$_{4}$ nanoparticles, which implies that
coating nanoparticles with SiO$_{2}$ can effectively protect the nanoparticles from oxidation or reduction.
} 
\end{figure}

Since about 80$\%$ of $\alpha$-Fe$_{2}$O$_{3}$ nanoparticles are not
reduced between 850 and 930 K due to the protection of SiO$_{2}$, we can determine the N\'eel temperature 
of the 60-nm $\alpha$-Fe$_{2}$O$_{3}$ nanoparticles 
by extrapolation of the warm-up magnetization data to higher temperatures. 
The solid line in Fig.~5 is a fit using the $M(T)$ curve of a bulk $\alpha$-Fe$_{2}$O$_{3}$ sample \cite{Di}, appropriately scaled to $T_{N}$ =
945 K. The inferred $T_{N}$ of 945 K for the 60-nm $\alpha$-Fe$_{2}$O$_{3}$ nanoparticles
is about 15 K lower than the bulk value of 960 K (Ref.~\cite{Di}).

\begin{figure}[htb]
     \vspace{-0.2cm}
    \includegraphics[height=6.4cm]{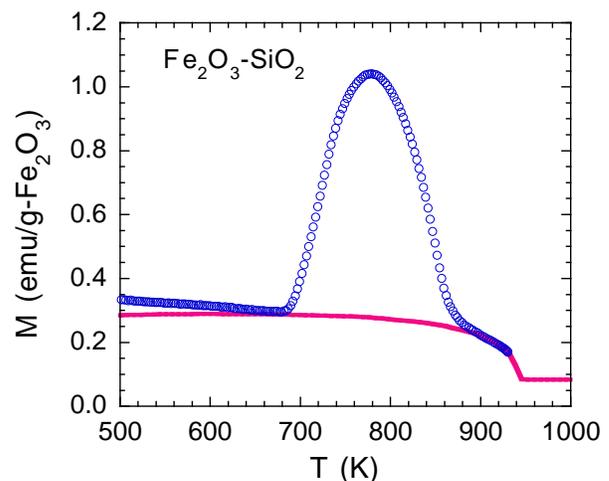}
     \vspace{-0.4cm}
 \caption[~]{Temperature dependence of the magnetization for the $\alpha$-Fe$_{2}$O$_{3}$ 
 nanoparticles coated with SiO$_{2}$. The solid line is a fit 
 using the $M(T)$ curve of a bulk $\alpha$-Fe$_{2}$O$_{3}$ sample \cite{Di}, appropriately scaled to $T_{N}$ =
945 K. The inferred $T_{N}$ for the $\alpha$-Fe$_{2}$O$_{3}$ nanoparticles
is about 15 K lower than the bulk value of 960 K.} 
\end{figure}

The reduction of the N\'eel temperature $T_{N}(d)$ of 
the nanoparticles from the bulk value $T_{N}(\infty)$ is due to finite-size
effects. The finite-size effects lead to a simple scaling
relationship~\cite{Fisher}:

\begin{equation}
    T_{N}(\infty) - T_{N}(d) = T_{N}(\infty)(d_{0}/d)^{1/\nu},
    \end{equation}
where $d_{0}$ is a microscopic length scale that is in the same
order of the lattice constant $a$ (Refs.~\cite{Fisher,Huang}).  The exponent 
$\nu$ is predicted to be 0.7048 for a pure three-dimensional Heisenberg 
magnet \cite{Chen} and 0.6417 for the Ising systems \cite{Fer}.  For
antiferromagnetic thin films of CoO, $d_{0}$ = 2 nm $\simeq$ 5$a$ and $\nu$ 
= 0.65 (Ref.~\cite{Amb95}). The measured exponent is in better agreement with the
theoretical prediction for the Ising systems~\cite{Chen}, which is
consistent with the fact that Co behaves more like an Ising than a
Heisenberg system \cite{Sal}. If we adopt $\nu$ = 0.65 and $d_{0}$ = 5$(a^{2}c)^{1/3}$ = 
3.5 nm for $\alpha$-Fe$_{2}$O$_{3}$ (where the lattice
constants $a$ = 0.50 nm and $c$ = 1.38 nm \cite{Di}), we 
calculate the $T_{N}$
reduction of 60-nm nanoparticles to be 12.0 K. If we use
$\nu$ = 0.70 and $d_{0}$ = 3.5 nm, we find the $T_{N}$ suppression of
16.5 K. Both values are in good
agreement with the measured one (15 K). 
                                                      
On the other hand, a $T_{N}$ reduction of about 100 K was reported
for single-crystal $\alpha$-Fe$_{2}$O$_{3}$
nanowire arrays with $d$ = 150 nm \cite{Di}. This reduction
seems too large to be compatible with Eq.~(1) which gives a
$T_{N}$ reduction of about 4.5 K for $d$ = 150 nm. Since we have shown
that $\alpha$-Fe$_{2}$O$_{3}$ nanoparticles are easy to reduce to the Fe$_{3}$O$_{4}$ phase in 
the high-vacuum environment,  the argon
environment for the magnetic measurement of the $\alpha$-Fe$_{2}$O$_{3}$
nanowire arrays \cite{Di} should also cause reduction of the $\alpha$-Fe$_{2}$O$_{3}$ to Fe$_{3}$O$_{4}$
phase. Therefore, the reported $T_{N}$ of 852 K for the nanowire arrays 
\cite{Di} is very likely to be associated with the ferrimagnetic transition of the 
converted Fe$_{3}$O$_{4}$ phase.

In summary, magnetic measurements up to 1000 K have been performed on hydrothermally 
 synthesized $\alpha$-Fe$_{2}$O$_{3}$ nanoparticles using a Quantum
 Design vibrating sample magnetometer. The high vacuum environment
 (1$\times$10$^{-5}$ torr) during the magnetic measurement up to 1000 K leads to a complete 
 reduction of $\alpha$-Fe$_{2}$O$_{3}$  to Fe$_{3}$O$_{4}$. This
 precludes the determination of the N\'eel temperature of the $\alpha$-Fe$_{2}$O$_{3}$
 nanoparticles. In contrast, coating
 $\alpha$-Fe$_{2}$O$_{3}$ nanoparticles with SiO$_{2}$ can effectively protect the nanoparticles
 from reduction, which allows us to determine the N\'eel temperature of the $\alpha$-Fe$_{2}$O$_{3}$
 nanoparticles for the first time. The N\'eel temperature of the 60-nm 
 $\alpha$-Fe$_{2}$O$_{3}$ nanoparticles is found to be 945 K, about 15 K below the 
 bulk value. The small reduction of the N\'eel temperature of the $\alpha$-Fe$_{2}$O$_{3}$ 
 nanoparticles is consistent with a finite-size scaling theory.

{\bf Acknowledgment:}
We would like to thank L. H. Meng for assistance in preparation of
Fe$_{3}$O$_{4}$-SiO$_{2}$ composites. This work was supported by the National Natural Science Foundation of China (10874095), 
the Science Foundation of China, Zhejiang (Y407267, 2009C31149), the Natural Science Foundation of Ningbo 
(2008B10051, 2009B21003), K. C. Wong Magna Foundation, and Y. G. Bao's Foundation. 

~\\
~\\
$^{a}$ wangjun2@nbu.edu.cn~\\
$^{b}$ gzhao2@calstatela.edu

\bibliographystyle{prsty}

\end{document}